\RequirePackage{fix-cm}
\documentclass[smallextended]{svjour3}       
\smartqed  
\usepackage{color}
\usepackage{graphicx}
\usepackage[justification=centering]{caption}
\begin{document}

\title{Same initial states attack in Yang et al.'s quantum private comparison protocol and the improvement
} \subtitle{}


\author{Wen-Jie Liu       \and
        Chao Liu          \and
        Zhi-Hao Liu       \and
        Jing-Fa Liu       \and
        Huan-Tong Geng    
}


\institute{
           W.-J. Liu  \and
           J.-F. Liu
           \at Jiangsu Engineering Center of Network Monitoring, Nanjing 210044, China
           \\
           \email{wenjiel@163.com}
           \and
           W.-J. Liu \and
           C. Liu \and
           J.-F. Liu \and
           H.-T. Geng
           \at School of Computer and Software, Nanjing University of Information Science \& Technology, Nanjing 210044, China
           \and
           Z.-H. Liu
           \at School of Computer Science and Engineering, Southeast University, Nanjing 211189, China
}

\date{Received: date / Accepted: date}

\maketitle

\begin{abstract}
In Yang \emph{et al.}'s literatures [J. Phys. A: Math. 42,
055305, 2009; J. Phys. A: Math. 43, 209801, 2010], a
quantum private comparison protocol based on Bell states and hash
function is proposed, which aims to securely compare the equality of
two participants' information with the help of a dishonest third
party (TP). However, this study will point out their protocol cannot
resist a special kind of attack, TP's same initial states attack,
which is presented in this paper. That is, the dishonest TP can
disturb the comparison result without being detected through
preparing the same initial states. Finally, a simple improvement is
given to avoid the attack.
\keywords{Quantum cryptography \and Quantum computation \and Quantum private comparison \and Same
initial states attack}
\end{abstract}

\section{Introduction}
\label{intro1}~~~~The principles of quantum mechanics, such as
no-cloning theorem, uncertainty principle, and \textcolor[rgb]{0,0,0.65}{entanglement characteristics}, provide
some interesting ways for cryptography communication and secure
computation. During the past thirty years, quantum communication has
developed in a variety of directions, including quantum key
distribution (QKD) [1, 2], quantum secret sharing (QSS) [3, 4],
quantum direct communication (QDC) [5-7], quantum teleportation (QT)
[8, 9], etc. On the other hand, secure multi-party computation (SMC)
has also been discussed in the quantum mechanism. Many special SMC
problems have been solved in quantum setting, for instance, quantum
private comparison (QPC) [10-19], quantum protocols for millionaire
problem [20, 21], quantum voting [22, 23] and quantum auctions [24,
25].

As an important branch and the foundation of quantum secure
multi-party computation (QSMC), QPC has attracted more and more
attention. The main goal of QPC is to compare the equality of secret
inputs between two participants without disclosing any information
about each other's secret content. The pioneering QPC protocol was
proposed by Yang \emph{et al.} [10] in 2009, and then in 2010, they revised
the protocol by removing the step (4) which is in fact unnecessary
[26]. \textcolor[rgb]{0,0,0.65}{Enlightened by Yang \emph{et al.}'s work, more QPC protocols
are proposed subsequently [11-18].}

However, while revisiting Yang \emph{et al.}'s literature [10] and its
corrigendum [26], we find there is a loophole undiscovered up to
date, and the dishonest TP can prepare the same initial states
instead of original random EPR pairs to confuse the two
participants' comparison result (we call it as TP's same initial
states attack). Under this kind of attack, TP is likely to make the
participants get the wrong result without being detected, which
directly results in the failure of private comparison. What's more,
TP can know the comparison result of the participants, which is
contrary to what they claimed: ``TP cannot learn any information
about the participants' respective secret inputs and even about the
comparison result".

The structure of this paper is organized as follows. At first, Yang
\emph{et al.}'s protocol is briefly reviewed in Section 2. In Section 3,
TP's same initial states attack is introduced and the loophole is
pointed out, and then a simple improvement is given as well.
Finally, a brief summary is concluded in Section 4.

\section{Review of Yang et al.'s protocol}
\label{sec:2}~~~~In Ref. [10], Yang \emph{et al.} presented a QPC protocol
based on Bell states and hash function, which aims to securely
compare the equality of two participants' secret inputs ($x$, $y$). It
should be noted that the third party (TP) is assumed to be
dishonest. However, they find the step (4) is in fact unnecessary,
and the protocol security will be ensured by the security check in
steps (5) and (6). In order to improve the efficiency of the
original protocol, Yang \emph{et al.} removed the first security check in
step (4) in Ref. [26]. Combining the thought of Ref. [10] and its
corrigendum [26], the main procedures can be described as follows
(also shown in Figure 1).

\textbf{Step 1.} Bob, Charlie and TP agree that these four unitary
operations $U_{i}(i=00,01,10,11)$ represent two-bit information
'00', '01', '10' and '11', respectively.
\begin{equation}\label{Eqau}
\left\{
\begin{array}{rcl}
& U_{00}=I=|0\rangle\langle0|+|1\rangle\langle1|,~~~\\
& U_{01}=\sigma_{z}=|0\rangle\langle0|-|1\rangle\langle1|,~\\
& U_{10}=\sigma_{x}=|1\rangle\langle0|+|0\rangle\langle1|,~\\
& U_{11}=i\sigma_{y}=|0\rangle\langle1|-|1\rangle\langle0|.
\end{array} \right.
\end{equation}
At the same time, Bob and Charlie share a secret hash function $H$ beforehand, the hash values of $x$, $y$ are $ H(x)=(x^{'}_{M-1},x^{'}_{M-2},...,x^{'}_{0})$, $H(y)=(y^{'}_{M-1},y^{'}_{M-2},...,y^{'}_{0})$, respectively, and $M$ denotes the length of the hash values.

\textbf{Step 2.} TP prepares $n(n>M/2)$  ERP pairs, each of which is
randomly one of the four Bell states
$|\Phi^{\pm}\rangle=\frac{1}{{\sqrt{2}}}(|00\rangle\pm\langle11|)$,
$|\Psi^\pm\rangle=\frac{1}{{\sqrt{2}}}(|01\rangle\pm\langle10|)$.
Later, TP divides these $n$ EPR pairs into two sequences $T_{B}$ and
$T_{C}$, which are composed of the first and the second particles of
each Bell state, respectively. Then TP randomly selects decoy
photons ${|0\rangle,|1\rangle,|+\rangle,|-\rangle}$ and inserts
these decoy photons into $T_{B}$ and $T_{C}$ at random positions,
respectively. Finally, TP sends the two sequences to Bob and
Charlie.

\textbf{Step 3.} Having received all particles, Bob and Charlie
check whether there is any eavesdropper in quantum channel by
measuring these inserted decoy photons. If there is no eavesdropper,
they continue to the next step; otherwise, they will terminate the
protocol.

\textbf{Step 4.} Bob and Charlie perform unitary operation on the
remaining particles according to $H(x)$ and $H(y)$. They perform $U_{B}$ and
$U_{C}$ on the ${(k+1)}$th photon, respectively.

\begin{equation}
U_{B}=\left\{
\begin{array}{rcl}
U_{x^{'}_{2k}x^{'}_{2k+1}}& & if~M~is~even\\
U_{x^{'}_{2k-1}x^{'}_{2k}}& & if~M~is~odd
\end{array} \right.
\end{equation}
\begin{equation}
U_{C}=\left\{
\begin{array}{rcl}
U_{y^{'}_{2k}y^{'}_{2k+1}}& & if~M~is~even\\
U_{y^{'}_{2k-1}y^{'}_{2k}}& & if~M~is~odd
\end{array} \right.
\end{equation}
where $k=0,1,\ldots|\frac{M-2}{2}|$ and $x^{'}_{2k}x^{'}_{2k+1}$,
$x^{'}_{2k-1}x^{'}_{2k}$; $y^{'}_{2k}y^{'}_{2k+1}$,
$y^{'}_{2k-1}y^{'}_{2k}$ are their hash values' bits of secret
inputs. In order to check whether TP will cheat in the following
announcement of his measurement outcomes, Bob and Charlie should
make some disarrangement operations as follows, firstly they
secretly generate a random number $l$ by using the QKD method, and
then insert the remaining intact EPR photons into the encoded photon
sequence at the positions determined by the value of $l$,
respectively. Besides, Bob and Charlie require TP to publish the
states of the remaining EPR pairs beforehand. For checking
eavesdropping in the Bob-TP and Charlie-TP quantum channels, Bob and
Charlie choose some decoy photons randomly from
$\{|0\rangle,|1\rangle,|+\rangle,|-\rangle\}$ and insert them into
EPR photon sequences at random positions. Finally, they send the
sequences back to TP.

\textbf{Step 5.} After received all particles, TP, Bob and Charlie
check the security of their quantum channels using the same method
as Step 3. If they confirm no eavesdropping, TP takes the Bell-basis
measurement on each two correlated photons received from Bob and
Charlie, records these measurement outcomes and publishes his
initial states of EPR pairs except for eavesdropping check. These
measurement outcomes are composed of two sets: the set of the
sampling EPR pairs' measurement outcomes $C$, and the set of the
encoding EPR pairs' measurement outcomes $M$. \textcolor[rgb]{0,0,0.65}{It should be noted
that} TP cannot know the information about which set each
measurement outcome of EPR pair belongs to. Bob and Charlie choose a
subset of the positions from one of the two sets randomly and ask TP
to publish the measurement outcomes at these positions. For the
positions chosen from $C$, \textcolor[rgb]{0,0,0.65}{if} the inconsistency rate between the
measurement outcomes \textcolor[rgb]{0,0,0.65}{and TP's beforehand announcements} is higher
than the threshold, Bob and Charlie know that TP is cheating and
abort the protocol. Otherwise, they continue to choose a position
subset randomly from one of the two sets and ask TP to public the
measurement outcomes. For the positions chosen from $M$, Bob and
Charlie can deduce the comparison result according to TP's
measurement outcomes and their initial states.

\begin{figure}
  \centering
    \includegraphics[width=3.5in]{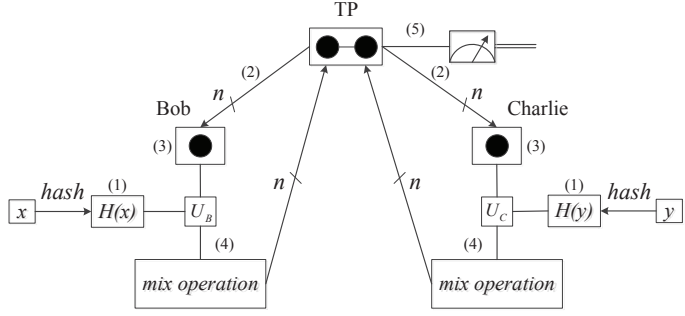}
  \centering
  \caption{The process of Yang \emph{et al.}'s QPC protocol based on Bell states and hash function.}      
\end{figure}

\textcolor[rgb]{0,0,0.65}{According to the above description}, it is not difficult to find Yang \emph{et al.}'s protocol has the following features:
\begin{enumerate}
  \item The protocol is based on EPR pairs and hash function. The secret inputs are encrypted by one-way hash function firstly, \textcolor[rgb]{0,0,0.65}{and then encoded into EPR pairs by using local unitary operations}.
  \item TP is dishonest, which is different from most of the other QPC protocols [11-18] that assume TP is semi-honest. In addition, TP is demanded to have no knowledge of two participants' private information and even the comparison result.
  \item The whole ERP pairs sequence is composed of two sets: the encoding set ($M$ is the corresponding measurement \textcolor[rgb]{0,0,0.65}{outcomes}) and the sampling set ($C$ is the measurement \textcolor[rgb]{0,0,0.65}{outcomes}). The first set is used to compare the two participants' secret inputs, while the other set is applied to check whether TP is honest. In addition, In order to prevent TP from knowing the positions of the sample set and further cheating the participants without being detected, the disarrangement operations on the EPR photons are indispensable in the process of Yang \emph{et al.}'s QPC protocol.
\end{enumerate}

In the protocol, quantum superdense coding [27] method is utilized
to conceal private information. Because TP \textcolor[rgb]{0,0,0.65}{needs} to publish both the
initial states and the final measurement outcomes of EPR pairs, the
participants can deduce the unitary operations that the counterparty
used (i.e., the corresponding hash value), but they cannot know the
actual bit value of the secret inputs due to the characteristic of
one-way hash function. It is ostensible that the scheme is secure,
but the security is mainly based on the classic cryptography, i.e.,
the one-way hash function, instead of the quantum cryptography
mechanism. Moreover, since TP is assumed to be dishonest, so he/she
can make any attacks. That is to say, TP is likely to launch the
same initial states attack to make the participants get wrong
comparison result without being detected, and more details will be
described in Section 3.

\section{TP's same initial states attack and the improvement}
\subsection{TP's same initial states attack on Yang et al.'s protocol}
\label{sec:3} ~~~~Suppose TP launches the same initial states attack
on Yang \emph{et al.}'s protocol, the detailed process is as follows: TP
firstly prepares the EPR pairs sequence in the same Bell state (e.g.
$|\Phi^+\rangle$) in Step 2, and the $|\Phi^+\rangle$ sequence is
divided into two sequences, $T_{B}$ and $T_{C}$, which will be sent
to Bob and Charlie, respectively. Secondly, TP will maliciously
publish that the final measurement outcomes are the same as their
initial states in Step 5. It is obvious that the disarrangement
operations in Step 4 will lose the original effect under this
special kind of attack. Since all the EPR pairs are the same, Bob
and Charlie cannot detect TP's cheating when he/she announces his/her
measurement outcomes. And there are two cases as follows:
\begin{enumerate}
  \item If the relation of Bob's and Charlie's secret inputs is $x\neq y$, but TP maliciously publishes that the measurement outcomes (including $C$ and $M$) are the same as the initial states he/she prepared (i.e., $|\Phi^+\rangle$) in Step 2. Since the measurement outcomes of $C$ are the same as the initial states, Bob and Charlie cannot perceive TP's cheating behavior. Hence, on the basis of the relation between TP's measurement outcomes and their initial states, they will mistakenly believe that their secret messages satisfy $x=y$, which means they get a wrong comparison result. What's more, TP will know the correct comparison result because of knowing that the measurement outcomes are different from the initial states.
  \item If \textcolor[rgb]{0,0,0.65}{$x=y$}, and TP still publishes that the measurement outcomes are the same as the initial states, in this case, Bob and Charlie can gain the correct comparison result (i.e., \textcolor[rgb]{0,0,0.65}{$x=y$}), but the result is also known by TP.
\end{enumerate}

As discussed above, if TP launches the same initial states attack on
Yang \emph{et al.}'s protocol, whatever \textcolor[rgb]{0,0,0.65}{$x=y$} or not, he/she will
always know the correct comparison result. What's more, if $x\neq
y$, Bob and Charlie will get wrong comparison result without being
detected.

\subsection{The improvement of Yang et al.'s protocol}
~~~~To avoid TP's same initial states attack, we can add a random
unitary operation procedure to fix the loophole of Yang \emph{et al.}'s
protocol. The detailed process is as follows: In Step 4, before Bob
and Charlie perform the disarrangement operations, they need to
randomly choose one of four unitary operations $U_{i}$ (see Equation
1) to perform on the sampling EPR pair photons, respectively. That
means, the final states of these sampling EPR pairs are not always
the same as their initial states. In addition, in order to check
whether TP is cheating or not, Bob and Charlie are required to
publish \textcolor[rgb]{0,0,0.65}{the unitary operations on the sampling
photons} before TP announces the measurement outcomes in Step 5.

In the improved protocol, even if TP tries to launch the same
initial states attack, the cheating will be detected because that
the inconsistent rate among the random unitary operations, TP's
announcement of $C$ and their initial EPR pair states exceeds a
rational threshold. In other words, the improved protocol is also
secure against the other conventional attacks presented in Ref.
[10].

\section{Conclusion}
\label{sec:4}~~~~In this paper, by revisiting and analysing Yang \emph{et al.}'s QPC
protocol, we firstly point out there is a security loophole in it.
More specifically, TP may launch the same initial states attack to
confuse the two participants' comparison results, what's more, the
result will be revealed to TP. In order to fix the loophole, a
simple improvement is proposed by adding a random unitary operation
procedure before the disarrangement operations. Analysis shows the
improved protocol can resist TP's same initial states attack.

\begin{acknowledgements}
This work is supported by China National Nature Science Foundation
(Grant Nos. 61103235, 61070240 and 61170321), A Project Funded by
the Priority Academic Program Development of Jiangsu Higher
Education Institutions, and Natural Science Foundation of Jiangsu
Province, China (BK2010570).
\end{acknowledgements}



\end{document}